\documentclass[
aps,%
12pt,%
final,%
notitlepage,%
oneside,%
onecolumn,%
nobibnotes,%
nofootinbib,%
superscriptaddress,%
noshowpacs,%
centertags]%
{revtex4}
\usepackage{graphicx}
\usepackage{mathtools}
\begin{document}
\selectlanguage{english}

\title{Finding Strangelets in Cosmic Rays from HESS J1731-347,\\ a Possible Strange Quark Star\\ Using the Cherenkov Telescope Array Observatory}

\author{\firstname{C.~R.}~\surname{Das}}

\email{das@theor.jinr.ru}

\affiliation{The Bogoliubov Laboratory of Theoretical Physics, International Intergovernmental Scientific Research Organization, Joint Institute for Nuclear Research, Dubna, Russia}

\date{Received August 11, 2025; revised August 11, 2025; accepted August 11, 2025}

\begin{abstract}

\noindent {\bf Abstract---}The hypothesis that supernova remnants are key sources of Galactic cosmic rays gains support from evidence that HESS J1731-347 one of the few Galactic objects capable of accelerating hadronic cosmic rays to TeV energies may harbor an exotic strange quark star rather than a conventional neutron star. This conclusion stems from its unusually low mass and compact radius, which challenge standard neutron star models. If confirmed, such a quark star could generate cosmic rays through the transition from the two-flavor color-superconducting (2SC) phase to the color-flavor-locked (CFL) phase, potentially releasing strangelets, hypothetical strange quark matter (SQM) particles. Detecting these strangelets in cosmic rays would provide groundbreaking evidence for quark matter. The future Cherenkov Telescope Array (CTA), with its unmatched sensitivity and spectral resolution in the very-high-energy (VHE) gamma-ray regime, is uniquely positioned to search for their annihilation or decay signatures. We analyze theoretical predictions for these gamma-ray signals and evaluate CTA's potential to detect or constrain them. Additionally, we present an in-depth assessment of CTA observations of HESS J1731-347, focusing on spectral features that could confirm strangelet production. A positive detection would not only validate the existence of strange quark stars but also establish a direct link between quark matter and cosmic-ray acceleration, reshaping our understanding of compact objects and high-energy astrophysics.
  
\end{abstract}

\maketitle

\section{Introduction}

Strange Quark Matter (SQM), often referred to as strangelets or nuclearites, is a hypothetical state of matter composed of roughly equal numbers of up, down, and strange quarks. Numerous theoretical studies and simulations suggest that SQM may represent the fundamental state of matter, potentially more stable than ordinary hadrons such as protons and neutrons. The inclusion of strange quarks introduces a third Fermi energy level, which can lower the energy per nucleon, thereby stabilizing this exotic form of matter. Intriguingly, SQM could account for a portion or even the entirety of the non-baryonic dark matter inferred from cosmological observations, without requiring new fundamental fields beyond the Standard Model of particle physics. Moreover, since SQM would have formed prior to primordial nucleosynthesis, its existence would not disrupt standard cosmological calculations.

If SQM is indeed the ground state of matter, it could profoundly influence the structure of compact astrophysical objects. Stars composed entirely of SQM, novel types of white dwarfs, or neutron stars with complex interiors characterized by diverse hadronic compositions and density profiles might exist. Additionally, strong transient astrophysical emissions could arise from internal phase transitions within these compact objects, offering potential observational signatures.

Recent measurements of the compact object HESS J1731-347, with a mass of $M = 0.77^{+0.20}_{-0.17} M_\odot$ and a radius of $R = 10.4^{+0.86}_{-0.78}$ km, place it among the lightest and smallest compact objects ever observed, pending further validation. These properties suggest that HESS J1731-347 could be an unusually light neutron star, a peculiar quark star, or a hybrid star undergoing an early deconfinement phase transition. Such characteristics raise fundamental questions about its nature and open the door to various theoretical interpretations. The emergence of strange quarks, combined with vector interactions, chiral symmetry restoration, and deconfinement phase transitions, may create a metastable state conducive to strangelet formation.

Should strangelets undergo disintegration or annihilation, they could produce distinctive spectral features in the gamma-ray spectrum. The Cherenkov Telescope Array (CTA) Observatory, with its exceptional sensitivity and energy resolution, is ideally suited to search for these signals. As the most advanced facility for gamma-ray astronomy, the CTA is designed to detect very-high-energy (VHE) gamma rays in the energy range of 20 GeV to over 300 TeV. It can probe extreme astrophysical environments, including active galactic nuclei, supernova remnants, and potentially exotic phenomena involving strangelets or dark matter. With superior sensitivity, resolution, and sky coverage compared to existing instruments like MAGIC, VERITAS, and H.E.S.S., the CTA is poised to investigate high-energy astrophysical phenomena, such as possible signatures of strangelets in the context of HESS J1731-347. This document summarizes the theoretical predictions for strangelet-related gamma-ray signatures, their detectability with the CTA, and their implications for constraining the properties of strangelets and HESS J1731-347.

\section{Scientific Goals}

\subsection{Strange Quark Matter and Current Searches}

The concept of SQM was first proposed in the 1980s as a distinct state of hadronic matter, coexisting alongside conventional nuclear matter \cite{PhysRevD.4.1601,PhysRevD.30.272,PhysRevD.30.2379}. The mass of strangelets within compact stars can range from a minimum stable mass of approximately $10s A$ \cite{PhysRevD.9.3471} to as high as $A \simeq 10^{57}$, equivalent to a solar mass of approximately $2 \times 10^{30}$ kg \cite{haensel1986strange,PhysRevD.71.014026}. SQM could have formed in the early universe shortly after the Big Bang \cite{PhysRevD.9.3471}, potentially contributing to baryonic dark matter \cite{PhysRevD.90.045010}. It may reside within the cores of neutron stars or exist as ``strange quark stars,'' which could be either purely composed of quarks \cite{haensel1986strange,alcock1986strange} or a mixture of hadrons and quarks \cite{drago2016scenario1,drago2016scenario2,PhysRevD.89.043014}. Strangelets could also be accelerated in the ergosphere of black holes \cite{BIANCHI2020115010} or produced in collisions between binary neutron stars \cite{PhysRevD.106.103032,Drago_2018,Wiktorowicz_2017}, potentially reaching Earth \cite{de1984nuclearites}.

Despite extensive searches, no definitive evidence for strangelets has been found, though experimental efforts have established upper limits on their production. The CERN NA52 experiment, which investigated strangelet production in Pb--Pb collisions at 158 GeV/$c$ per nucleon, found no evidence and set upper limits on production yields for strangelets with mass-to-charge ratios up to $m/|Z| \approx 60$ GeV/$c^2$ \cite{PhysRevLett.76.3907}. Similarly, the PAMELA experiment, which searched for cosmic-ray strangelets from 2006 to 2009, detected no candidates and established an upper limit on the strangelet flux for particles with charges $1 \leq Z \leq 8$ and baryon numbers $4 \leq A \leq 1.2 \times 10^5$ \cite{PhysRevD.71.083008}.

\subsection{HESS J1731-347 and Strong First-Order (Deconfinement) Phase Transition}

The High Energy Stereoscopic System (H.E.S.S.), an array of imaging atmospheric Cherenkov telescopes in Namibia, discovered the VHE gamma-ray source HESS J1731-347. Located in the constellation Scorpius near the Galactic plane, it has celestial coordinates of right ascension (RA) 17 h 31 m, declination (Dec) $-34^\circ 47'$, and galactic coordinates of longitude $l \simeq 353.6^\circ$ (near the galactic center) and latitude $b \simeq -0.7^\circ$ (slightly below the galactic plane). Its spatial coincidence with the shell-type supernova remnant (SNR) G353.6-0.7 suggests a potential association. The gamma-ray emission, observed at energies above 100 GeV, may result from accelerated cosmic rays interacting with surrounding material through processes such as pion decay or inverse Compton scattering. While its precise distance remains uncertain, estimates place it at approximately 10400 light-years (or $\sim$3.2 kiloparsecs) if associated with SNR G353.6-0.7 \cite{Sagun_2023,horvath2023light,char2024compact,zhang2024can}.

The conditions under which quark matter forms and the density at which deconfinement phase transitions occur remain unclear. Combined analyses of multimessenger constraints and elliptic flow in heavy-ion collisions suggest that strongly interacting matter softens at high densities, potentially indicating a transition to quark-gluon plasma \cite{annala2022multimessenger,huth2022constraining}. By correlating the measured mass, radius, and surface temperature of HESS J1731-347 with theoretical models of strongly interacting matter and color superconductivity at high densities, we can better understand its nature.

Recent evidence challenges the occurrence of a strong first-order deconfinement phase transition in neutron stars, particularly due to nontrivial phenomena associated with such transitions \cite{PhysRevD.110.123009,PhysRevD.108.094014}. The exceptionally low mass and radius of HESS J1731-347 \cite{doroshenko2022strangely} have sparked skepticism within the neutron star community, as state-of-the-art simulations indicate that the lightest neutron stars formed via supernova explosions have masses of at least $1.1 M_\odot$. No known scenario explains the existence of a lighter neutron star. For HESS J1731-347 to exhibit such properties, an early deconfinement phase transition would be required, characterized by a large branching ratio of scalar meson decay, $\eta_D \gtrsim 1.1$, in the color-flavor-locked (CFL) phase of color superconductivity, where $\eta_D = G_D / G_S$, with $G_S = 9.92$ GeV$^{-2}$ as the coupling strength in the scalar meson channel and $G_D$ as the coupling strength of a specific decay mode. This condition is necessary to achieve neutron star radii at or below those typical of binary radio pulsars ($\lesssim$12 km) \cite{ayriyan2025bayesian}.

Alternatively, HESS J1731-347 could be a hadron star (HS) or strange star (SS) with an early deconfinement phase transition occurring below twice nuclear saturation density. Such a star would possess a significant quark--gluon plasma core, potentially leading to rapid cooling via the direct Urca (DU) process involving active quarks. However, the two-color superconducting (2SC) phase, characterized by the absence of strange quarks and the formation of a {\it u-d} diquark condensate in specific color directions, suppresses rapid cooling and aligns with the observed surface temperature, as derived from a self-consistent calculation of the quark pairing gap within a chirally symmetric relativistic density functional (RDF) model \cite{Sagun_2023}.

Quark matter can be categorized based on the presence of strange quarks, with the 2SC phase defined by the absence of strange quarks and the CFL phase characterized by Cooper pairs where color and flavor properties are correlated in a one-to-one correspondence among three colors and three flavors \cite{ALFORD2009385c}. Additional quark pairing configurations, such as gapless 2SC, crystalline color superconductivity, or gapless CFL (gCFL), are also possible \cite{BUBALLA2005205}. The observed surface temperature of HESS J1731-347 aligns well with the 2SC phase and a first-order deconfinement phase transition, negating the need for the CFL phase, which would further suppress neutrino emission while still fitting the data \cite{Klahn_2015}. However, a hybrid equation of state (EoS) combining the MDI-APR1 (hadronic) and CFL (quark) EoS, with a phase transition modeled via Maxwell construction, can explain HESS J1731-347's properties using stable CFL quark matter. In contrast, hybrid models incorporating the CFL MIT Bag model fail to account for the masses of the heaviest observed pulsars \cite{Kourmpetis2025}. Studies further suggest that HESS J1731-347 is either a strange star or a hadron star \cite{rather2023quark}, with comparisons provided in Figs 1 and 2 of \cite{Sagun_2023}, Fig 3 of \cite{rather2023quark}, Fig 4 of \cite{Klahn_2015}, Fig 5 of \cite{sym16010111}, and Fig 12 of \cite{ayriyan2025bayesian}.

\subsection{Phase Transitions and Stranglet Formation}

Strangelets are stabilized by strong interactions and require a significant population of strange quarks, which may form in high-density environments where the strange quark mass is dynamically reduced. In the 2SC phase, up and down quarks of two colors form a BCS-like condensate with a gap in the quark spectrum, while strange quarks remain unpaired due to their higher mass, making strangelet formation less likely \cite{PhysRevC.87.045208,Shovkovy2012}. A transition to the CFL phase or unpaired quark matter with a reduced strange quark mass could facilitate strangelet formation. Magnetic fields may also influence phase stability, potentially affecting strangelet production \cite{Shovkovy2012}.

The CFL phase, occurring at higher densities, involves all three flavors (up, down, and strange) forming a highly symmetric superfluid state where all fermionic modes are gapped. This phase's flavor-symmetric pairing and lower free energy enhance its stability, making it conducive to strangelet formation \cite{RevModPhys.80.1455}. Chiral symmetry restoration at high density reduces the strange quark mass, further promoting strangelet formation within the CFL condensate \cite{Rapp1998}. Mixed phases between the 2SC and CFL states may serve as nucleation sites for strangelets, though their stability is uncertain due to surface and Coulomb effects \cite{Schmitt2010}.

\subsection{Stranglet Formation in Astrophysical and Experimental Contexts}

In neutron stars, such as HESS J1731-347, the CFL phase's stability under neutrality constraints supports strangelet formation. In heavy-ion collisions, transient 2SC or CFL phases may produce strangelets, potentially detectable through negative pion radiation or proton excess \cite{Rapp1998,Kourmpetis2025}. The CFL phase is more favorable for strangelet formation than the 2SC phase due to its inclusion of strange quarks and lower free energy. However, instabilities in mixed phases and external factors like magnetic fields introduce uncertainties. Observational signatures, such as r-mode damping in neutron stars or particle production in collisions, could provide evidence for strangelets \cite{Kourmpetis2025,Shovkovy2012}.

\section{Future Experimental Goals}

\subsection{The CTA Observatory for Strangelet Searches}

The CTA is a next-generation ground-based observatory designed to detect VHE gamma rays from 20 GeV to over 300 TeV, establishing it as the most advanced facility for gamma-ray astronomy \cite{Abdalla_2021}. The CTA aims to explore high-energy astrophysical phenomena, including cosmic ray acceleration, dark matter, and exotic particles like strangelets, particularly in the context of HESS J1731-347 within the TeV gamma-ray range (0.1--100 TeV) \cite{abramowski2011new,10.1063/1.4968930}. Unlike optical telescopes, the CTA detects gamma rays indirectly by observing Cherenkov radiation, blue light flashes produced when VHE gamma rays create particle showers in Earth's atmosphere \cite{cerenkov1934visible}. Comprising two arrays for full-sky coverage, the Southern Array (CTA-South) at Paranal Observatory, Chile (2600 m altitude), is optimized for galactic sources. The 1600-hour Galactic Plane Survey (GPS) will map TeV sources, including HESS J1731-347 \cite{sousa2025prospects}.

The process $SS \to \gamma\gamma$ describes a theoretical scenario in which two strangelets ($S$), hypothetical particles made of SQM, annihilate to produce two gamma-ray photons. The energy of each gamma-ray photon is given by $E_\gamma = m_S c^2$, where $m_S$ is the rest mass of the strangelet and $c$ is the speed of light. This detection is facilitated by observing Cherenkov radiation produced by air showers in Earth's atmosphere \cite{ACHARYA20133,BERNLOHR2013171}.

\subsection{CTA Spectral Line Detection for HESS J1731-347}

The CTA offers unparalleled sensitivity to gamma-ray lines in the 0.1--10 TeV range, where its effective area is optimized for detecting the gamma-ray flux of HESS J1731-347. The differential flux at 1 TeV is $\frac{dN}{dE} = (3.0 \pm 0.6_{\rm stat} \pm 0.7_{\rm syst}) \times 10^{-12}$ cm$^{-2}$ s$^{-1}$ TeV$^{-1}$, with an integral flux above 1 TeV of approximately $3.5 \times 10^{-12}$ cm$^{-2}$ s$^{-1}$, roughly 3\% of the Crab Nebula's flux \cite{doroshenko2023expansion,Guo_2018}. Strangelets from HESS J1731-347 could produce line-like features through interactions with the interstellar medium (ISM) or exotic decay processes \cite{10.1007/BFb0107314,PhysRevD.106.103032}. With an energy resolution of approximately 10\% at 1 TeV, the CTA can detect lines with fluxes as low as $\sim$10$^{-13}$ ph/cm$^2$/s after 100 hours of observation, compared to H.E.S.S.'s limit of $\sim$10$^{-12}$ ph/cm$^2$/s \cite{Abdalla_2021}. For HESS J1731-347, with a flux of $\sim$10$^{-12}$ ph/cm$^2$/s at 1 TeV, the CTA can probe contributions below 0.1\% of the continuum flux. By fitting the spectrum to a power law ($dN/dE \propto E^{-2.3}$) and searching for Gaussian peaks, the strangelet flux can be constrained as follows:

\begin{equation}
F_s < \frac{F_\gamma^\text{line}}{n_H \cdot \sigma_s} \approx \frac{10^{-13}}{100 \cdot 4 \times 10^{-26}} \approx 2.5 \times 10^{-11} \, \text{ph/cm}^2/\text{s/sr}.
\end{equation}

Where $n_H$ is the average column density, $\sigma_s$ is the cross-section, and $F_\gamma^\text{line}$ is the line photon flux. With an angular resolution of approximately 1 arcminute, the CTA can map HESS J1731-347's morphology, correlating emission with ISM gas or neutron star remnants. Localized excesses could indicate strangelet production \cite{10.1007/BFb0107314}. The GPS, with $\sim$100 hours of observation, achieves sensitivity to faint signals (<$10^{-13}$ ph/cm$^2$/s), enabling tests of strangelet stability \cite{sousa2025prospects}.

\subsection{CTA Simulations for HESS J1731-347}

No studies have explicitly modeled strangelet spectral lines for the CTA in the context of HESS J1731-347 \cite{PhysRevD.81.024012,10.1007/BFb0107314}, as proton-driven $\pi^0$ hadronic gamma-ray emission dominates the observed spectrum. However, the CTA's Key Science Projects (KSPs) for dark matter and axion-like particles provide a framework for simulating strangelet-ISM interactions using the Gamma-ray Astronomy Science Analysis Software (\texttt{ctools}) \cite{jurgen_knodlseder_2021_4727876} based on \texttt{GammaLib} (A versatile toolbox for scientific analysis of astronomical gamma-ray data) \cite{jurgen_knodlseder_2021_4727871}, and \texttt{Gammapy} (A Python package for gamma-ray astronomy) \cite{donath2023gammapy} to predict spectral signatures \cite{Abdalla_2021}. To demonstrate the CTA's potential, we present a chart illustrating the upper limit on a hypothetical strangelet-induced spectral line flux versus gamma-ray energy (0.1--10 TeV), compared to H.E.S.S. limits and HESS J1731-347's continuum flux.

\begin{figure*}[t!]
\setcaptionmargin{5mm}
\onelinecaptionsfalse
\includegraphics{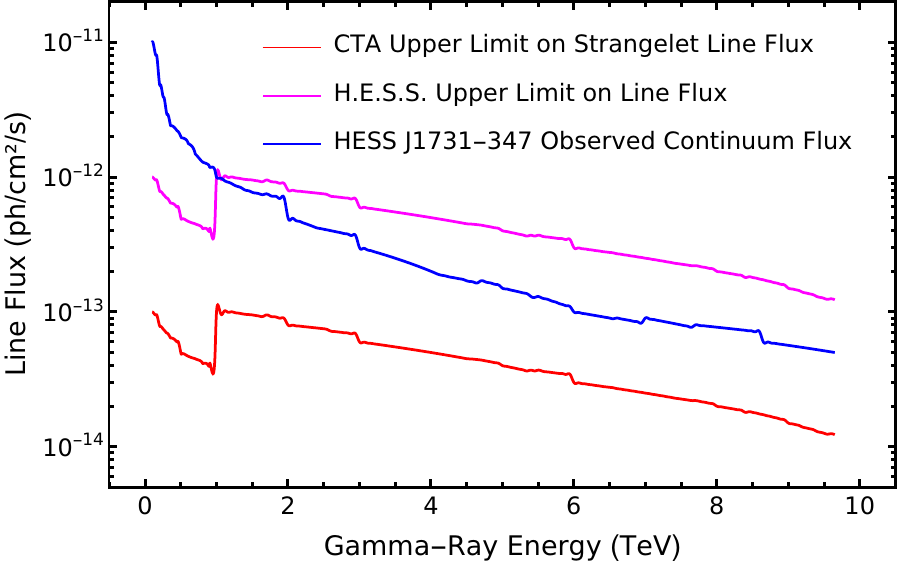}
\captionstyle{normal}
\caption{CTA limitations for strangelet spectral line flux (0.1--10 TeV). The plot depicts the upper CTA limit of strangelet-induced line flux (red), H.E.S.S. limit (magenta), and observed continuum flux (blue with error-band) for HESS J1731-347, with a logarithmic {\it y}-axis for flux (ph/cm$^2$/s).}\label{Figure: StrangeletFluxPlot}
\end{figure*}

\begin{figure*}[t!]
\setcaptionmargin{5mm}
\onelinecaptionsfalse
\includegraphics{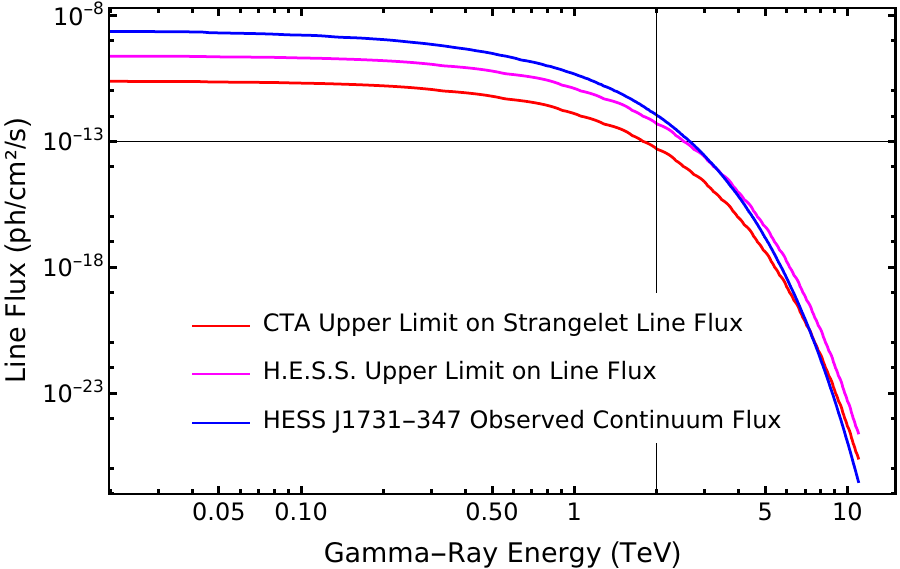}
\captionstyle{normal}
\caption{Similar to Fig \ref{Figure: StrangeletFluxPlot}, with a logarithmic {\it x}-axis, the CTA limit line at $\sim$10$^{-13}$ ph/cm$^2$/s, and enhanced Monte Carlo simulation. Simulation results for CTA limits over $\lesssim$10$^{-13}$ ph/cm$^2$/s beyond 2 TeV are unsatisfactory, as the HESS J1731-347 cutoff energy began at about 7--10 TeV.}\label{Figure: StrangeletFluxPlotExtend}
\end{figure*}

The Figs \ref{Figure: StrangeletFluxPlot} and \ref{Figure: StrangeletFluxPlotExtend}'s {\it x}-axis spans 0.1--10 TeV, where the CTA's Medium-Sized Telescopes (MSTs) and Small-Sized Telescopes (SSTs) are most sensitive. The {\it y}-axis, on a logarithmic scale, represents line flux (ph/cm$^2$/s), with the CTA's limit at $\sim$10$^{-13}$ ph/cm$^2$/s, reflecting its tenfold improvement over H.E.S.S.'s $\sim$10$^{-12}$ ph/cm$^2$/s. The continuum flux follows a power law ($dN/dE \propto E^{-2.3}$) \cite{10.1063/1.4968930}. Theoretical constraints indicate a line flux for HESS J1731-347 of <$10^{-13}$ ph/cm$^2$/s at 1 TeV, with a strangelet flux of $F_s < 2.5 \times 10^{-11}$ ph/cm$^2$/s/sr. The number density is estimated at <$3 \times 10^{-22}$ cm$^{-3}$ for strangelets with a total mass of <$10^{-17} M_\odot$ and a production rate of <$10^{-12}$ strangelets/s \cite{PhysRevD.81.024012}.

\section{Conclusions}

The 2SC phase, characterized by paired up and down quarks, is a superconducting state that may exist in the dense cores of neutron stars. The CFL phase, involving up, down, and strange quarks, exhibits enhanced flavor symmetry and stability, associated with SQM. The transition from the 2SC to the CFL phase can promote strangelet nucleation, potentially transforming a neutron star into a strange quark star, with stability influenced by magnetic fields, chiral symmetry restoration, and surface/Coulomb effects.

The CTA is optimized to detect VHE gamma rays, ranging from tens of GeV to hundreds of TeV. In the context of the compact object HESS J1731-347, which may harbor SQM in a strange star scenario, the CTA has the potential to detect or constrain gamma-ray lines resulting from strangelet annihilation. Additionally, the CTA's sensitivity to such annihilation signatures enables it to set competitive limits in dark matter searches, as these signatures are analogous to those expected from dark matter particle annihilation.

The CTA's advanced sensitivity, resolution, and sky coverage make it an ideal instrument for constraining strangelet spectral lines in HESS J1731-347, potentially setting stringent limits on flux and number density. The Galactic Plane Survey and multi-messenger approaches will further enhance these searches, necessitating new theoretical models to predict specific strangelet signatures and advance our understanding of exotic matter in the universe.

\bibliography{crdas}
\end{document}